# Controlling spin-orbit torque by polarization field in multiferroic BiFeO$_3$ based heterostructures

P. F. Liu [1], J. Miao [1] [a], Q. Liu [1], Z. D. Xu [2], Z. Y. Ren [1], K. K. Meng [1], Y. Wu [1], J. K. Chen [1], X. G. Xu [1] and Y. Jiang [1] [b]

[1] Beijing Advanced Innovation Center for Materials Genome Engineering, School of Materials Science and Engineering, University of Science and Technology Beijing, Beijing 100083, China

[2] Department of Physics, Southern University of Science and Technology, Shenzhen 518055, Guangdong, China

**Abstract**

In the last few years, some ideas of electric manipulations in ferromagnetic heterostructures have been proposed for developing next generation spintronic devices. Among them, the magnetization switching driven by spin-orbit torque (SOT) is being intensely pursued. Especially, how to control the switching current density, which is expected to enrich device functionalities, has aroused much interest among researchers all over the world. In this paper, a novel method to adjust the switching current is proposed, and the BiFeO$_3$ (BFO) based heterostructures with opposite spontaneous polarizations fields show huge changes in both perpendicular magnetic anisotropy and the SOT-induced magnetization switching. The damping-like torques were estimated by using harmonic Hall voltage measurement, and the variation of effective spin Hall angles for the heterostructures with opposite polarizations was calculated to be 272%. At the end of this paper, we have also demonstrated the possible applications of our structure in memory and reconfigurable logic devices.

[a] electronic mail: j.miao@ustb.edu.cn

[b] electronic mail: yjiang@ustb.edu.cn

The ability to manipulate magnetic states in ferromagnetic (FM) thin films has been widely studied for device applications such as nonvolatile magnetic memory. [1-2] Recently, it has been found that the combination of spin-transfer torque [3-4] and spin-orbit coupling (SOC) leads to a new type of torque, i.e. spin-orbit torque (SOT). [5-8] The advantages of SOT-induced magnetization switching are generally fast reaction, reliable operation, and feasible integration. [9] The origin of SOT arises from the bulk spin Hall effect (SHE) [10-12] in materials with strong SOC such as heavy metals (HMs) and/or the interfacial Rashba effect [13-14] at the interfaces. Especially in ferromagnet (FM)/HM bilayers, SHE can produce a damping-like torque (DLT), which has an anti-damping form, while the interfacial Rashba effect gives rise to a field-like torque (FLT). It is a general consensus that DLT is responsible for the magnetization switching of FM. [15] A problem of the SOT-induced magnetization switching is that it can only work under an external magnetic field for symmetric structures. The field is collinear with the input current in order to break the symmetry. Recently, two effective ways have been proposed to realize the SOT-induced magnetization switching in the absence of magnetic field, introducing a lateral structural asymmetry [16] and using antiferromagnet/FM bilayers exhibiting exchange-bias. [17-19] There have been some reports on the manipulation of SOT. For example, Wang et al. used an heterostructure of FM/HM/Pb $(Mg_{1/3}Nb_{2/3})_{0.72}Ti_{0.28}O_3$ (PMN-PT) [20] to realize the electric-field control of SOT, and Qiu et al. proposed that different oxygen levels in FM layer can largely influence SOT. [21]

$BiFeO_3$ (BFO) is a multiferroic material, whose rhombohedral structure gives rise to three types of domain walls including 71°, 109° and 180°. [22-25] For its ferroelectricity at room temperature, BFO has been widely studied. [26-28] However till now there is no report on the effect of BFO on SOT. An antisymmetric Dzyaloshinskii-Moriya interaction (DMI) in BFO gives rise to a weak ferromagnetic moment, [29,30] which is usually used for magnetoelectric coupling and may induce the field-free SOT switching. However, the DMI coupling can only occur in some special situations [30,31]. On the other hand, in the BFO films with different polarization fields, the oxygen migration may produce different distributions of oxygen vacancies. [32,33] If we deposit a FM/HM structure on a BFO film, the different polarization fields of BFO should affect SOT through the indirect control of the oxygen level in the FM layer [21], which is different from previous studies on PMN-PT. [20,34]

In this paper, we have studied the perpendicular magnetic anisotropy (PMA) and SOT-induced switching behaviors in a heterostructure of BFO/$Al_2O_3$/Pt/Co/Pt/Ta. Interestingly, the SOT-switching current density has been changed by 300% under the different directions of spontaneous polarization fields of the BFO film. Therefore, our study provides a method to control SOT through the mediation of polarization fields and oxygen vacancies distribution.

Two heterostructures of $SrTiO_3$ (STO)/BFO (50 nm)/$Al_2O_3$ (1 nm)/Pt (1 nm)/Co (1 nm)/Pt (1 nm)/Ta (3 nm) (BFO hetero) and STO/$SrRuO_3$ (SRO,10 nm)/ BFO (50 nm)/Al2O3(1 nm)/Pt (1 nm)/Co (1 nm)/Pt (1 nm)/Ta (3 nm) (S-BFO hetero) were prepared. First, the BFO and SRO/BFO films were separately deposited on Ti-terminated STO (100) substrates by pulsed laser deposition (PLD). Then the multilayers of Al2O3 (1 nm)/Pt (1 nm)/Co (0.7 nm)/Pt (1 nm)/Ta (3 nm) were grown on the BFO films by magnetic sputtering.

The growth temperatures of BFO and SRO are 650 ℃ and 680 ℃ respectively.

The SOT-induced magnetization switching behaviors were measured by Keithley 6221 AC current source and Keithley 2182. For harmonic measurement, a Keithley 6221 provided a small-amplitude sinusoidal AC current, while SR 830 was used for measuring harmonic voltages. The Atomic force microscope (AFM) image of the surface and the X-Ray Diffraction (XRD) pattern of the SRO (10 nm) /BFO (50 nm) bilayer are shown in Figure 1(a) and (b) respectively. The growth of the BFO film is layer by layer with the smoothness $R_q$ of about 0.2 nm. The diffraction peaks of (001) and (002) are strong for the BFO film which justifies the epitaxial growth of it. The multilayers were then micro-fabricated into Hall bars by electron beam lithography, ion-milling and lift-off process. More details of our fabrication process are shown in Supporting Information S1. The schematic diagram of the Hall bar structure is shown in Figure 1(c). The Hall bar (100 μm×60 μm) is surrounded by four Pt electrodes. The widths of the Hall bar are 20 μm (x-axis) and 10 μm (y-axis). The lattice parameters of the BFO film were obtained through the Reciprocal Space Mapping (RSM) test of the SRO/BFO bilayer in (103) orientation shown in Figure 1(d), ensuring that $a$ is about 0.393 nm, $c$ is about 0.406 nm, and therefore $c/a$ is 1.033nm.

According to Yu et al.,[32] the spontaneous polarization can be adjusted via inserting a buffer layer between STO and BFO. As shown in Figure S2(a) and (b) in the Supporting Information, the polarization field in BFO hetero is in up-state, while the one in S-BFO hetero is in down-state. Considering oxygen vacancies are usually formed at the side with negative polarization bound charges, the top surface of BFO can be modulated into oxygen-rich or oxygen-poor states with different directions of spontaneous polarization fields [33,35,36]. The amount of oxygen vacancies at the surface of BFO in S-BFO hetero should be larger than that in BFO hetero. [37-39] We have measured the anomalous Hall effect (AHE) of BFO hetero and S-BFO hetero. The anomalous Hall resistance ($R_H$) of both samples exhibit good PMA, as shown in Figure S2(c) and (d). The coercive field ($H_c$) of BFO hetero (80 Oe) is much larger than that of S-BFO hetero (30 Oe), which is in consistent with their M-H loops shown in Figure S2(e) and (f).

The SOT-induced magnetization switching behaviors have been studied for the two samples. As shown in Figure 2(a)-(d), under different in-plane magnetic fields, square-shaped hysteresis loops of $R_H$ can be observed in both samples. The changes of $R_H$ are 3.4 Ω and 0.4 Ω for the two samples, in consistent with the AHE measurement in Figure S2, which demonstrates the deterministic magnetization switching driven by SOT. For both samples, while the direction of the magnetic field changes from the positive x-axis to the negative, the switching curves are converted from clockwise to counter clockwise. Meanwhile, the critical current (in each polarization state) does not change much under different magnetic fields. Surprisingly, it is demonstrated the critical current $I_c$ in BFO hetero (~ 13 mA, $J_c$=1.08×10$^7$ A cm$^{-2}$) is much smaller than the one in S-BFO hetero (~ 40 mA, $J_c$=3.33×10$^7$ A cm$^{-2}$). Here $I_c$ is defined as the current corresponding to $(R_+ + R_-)/2$, where $R_+$ and $R_-$ are the saturation resistances under positive and negative current, respectively. Note that the Joule heating effect also needs to be clarified that it can raise the temperature of the samples and has a little contribution (not dominate) on the measured R-I signals using pulse currents. [40]

In Figure 3, the first $V_\omega$ and second $V_{2\omega}$ harmonic Hall voltages have been measured for the two samples to detect DLTs with a magnetic field aligning x-axis. The measurement current is 2.1 mA with the frequency of 111.2 Hz. The harmonic voltages under different frequencies of measurement current for the two samples are given in Figure S3 and Figure S4. Because the measured signals for S-BFO hetero under small frequencies of current were too noisy to detect, we used larger frequencies for the measurements. For the two samples, both $V_\omega$ and $V_{2\omega}$ as a function of magnetic field H almost do not change with different detecting frequencies. The odd behaviors of $V_{2\omega}$ along the x-axis indicate the existence of DLTs in both samples, which are independent of the direction of magnetic field and the frequency of applied current.

FLTs were also detected in harmonic measurements under the applied magnetic field parallel to the y-axis (Figure S5). Because the SOT switching part of our heterostructures (Pt/Co/Pt) is symmetrical, the contribution of FLTs to SOT is very weak compared to that of DLTs. Therefore, the changes of the critical switching current for the two samples shown in Figure 2 should be due to DLTs, rather than FLTs.

In order to quantitatively determine the strength of spin orbit effective field for the samples in different polarization fields, the $V_\omega$ and $V_{2\omega}$ under small in-plane magnetic fields have been measured, with a sinusoidal AC current of 1 mA ($J_c$=8.3×10$^6$ A cm$^{-2}$) and an out-of-plane magnetization component $M_z < 0$ or $M_z>0$. Before the measurements, the samples were saturated with a large out-of-plane external field, and the samples shall remain saturated after the large field was removed. Considering the negligible FLTs in both samples, we only present the measured $V_\omega$ and $V_{2\omega}$ in Figure S6.

According to Ohno et al [41], the damping-like effective field $H_D$ can be calculated using Equation (1).

$$H_D = -2\frac{H_L \pm 2\varepsilon H_T}{1-4\varepsilon^2}, \tag{1}$$

where $\varepsilon$ is the ratio of the planar Hall effect (PHE) resistance to $R_H$ with the calculation method of $\varepsilon$ shown in Supporting Information S7, and the $\pm$ sign represents the magnetization pointing to $\pm Z$. And the amplitude $H_L$ and $H_T$ along a given in-plane field direction can be determined via Eq. (2).

$$H_{L(T)} = \frac{\partial V_{2\omega}/\partial H_{X(Y)}}{\partial^2 V_\omega/\partial H_{X(Y)}^2}, \tag{2}$$

where $H_{X(Y)}$ is the magnetic field applied along the x(y)-axis. Based on our measurements, the calculated $H_D$ of the sample BFO hetero in $P_{up}$ state （122.01 Oe per A cm$^{-2}$） is much larger than that of S-BFO hetero in $P_{down}$ state (44.77 Oe per A cm$^{-2}$). We have further estimated the effective spin Hall angles (SHA) of the two samples by using Eq. (3). [42]

$$\theta_{SH} = \frac{2|e|M_S t_F H_D}{\hbar |j|}, \tag{3}$$

where $e$ is the charge of an electron, $M_S$ the saturation magnetization of the Co layer, $j$ the current density, and $t_F$ the thickness of the Co layer. The calculated effective SHA of BFO hetero is about 2.7 times larger than that of S-BFO hetero.

Referring to the opposite spontaneous polarization fields in BFO hetero and S-BFO hetero, we infer the different PMA and SOT between them are due to the different number of oxygen vacancies on the top surface of BFO. The oxidation degree of the Co layers, which plays a key role in the PMA and SOT behaviors of the two samples, is very sensitive to the oxygen vacancies distribution in BFO. [37] As shown in Figure 4(a) and (b), the top surface of BFO can be adjusted into oxygen-vacancy-rich (BFO hetero) or oxygen-vacancy-poor states (S-BFO hetero) according to the direction of spontaneous polarization field. Meanwhile, the different oxygen vacancies can influence the oxygen level in the Co layer deposited on BFO (See the Supporting Information S8). The critical current of SOT switching ($J_C$) is determined by Eq. (4).

$$J_C = \frac{2e}{\hbar} \frac{M_s t_F}{\theta_{SH}} \left(\frac{H_K}{2} - \frac{H_X}{\sqrt{2}}\right). \qquad (4)$$

Considering the effective SHA of BFO hetero is about 2.7 times larger than that of S-BFO hetero and the effective magnetic anisotropy ($H_K$) of BFO hetero is almost the same as that of S-BFO hetero (See Supporting Information S9), $J_C$ in BFO hetero should be much smaller than that in S-BFO hetero, as shown in Figure 4(e) and (f). It should be noted that a high level of oxidation of FM layer does not always bring high $K_{eff}$, and $K_{eff}$ should be improved by proper oxidation which is dependent on structures. [16]

Our heterostructures have good potential to be used in future memory and programmable logic devices if the polarization field of the BFO layer can be directly changed. The data are stored by the magnetization of the Co film with up state (+$m_z$) for 1 and down state (-$m_z$) for 0. The writing process could be realized by setting up a proper initial situation, changing writing current and external magnetic field, as schematically demonstrated in Figure 5(a). As in Figure 5(b), for programmable logic operations, the two currents $I_A$ and $I_B$ functioning as two inputs are applied along the x-axis, and the switching currents in P$_{up}$ state and P$_{down}$ state are 13 mA and 40 mA respectively. We set that $I_A$ and $I_B$ are 24 mA and 5 mA for Inputs 1 and 0 respectively, and $I_C$ values are the switching currents in opposite polarization fields (13 mA and 40 mA), thus the final magnetization state of the Co layer is determined by the value of ($I_A+I_B-I_C$) and the polarity of $H_x$.

Importantly, at least five logic gates (AND, OR, NOT, NOR, and NAND) can be realized in a single BFO-based heterostructure. In Figure 5(b), for an "AND" gate, the polarization is in P$_{down}$ state with the threshold current $I_C$ of 40 mA and $H_X$ of +200 Oe. Referring to the works of Han et al. [8,43], before logic operations, the device is initialized to -$m_z$, and then $I_A$ and $I_B$ are chosen as 1 or 0. If $I_A$ or $I_B$ is applied with 0, they cannot change the initial magnetization into +$m_z$ (Output is 0), therefore the heterostructure can be converted to +$m_z$ (Output is 1) only when $I_A=I_B=1$. As for the "OR" gate, the polarization is turned to be in P$_{up}$ state with the threshold current $I_C$ of 13mA and $H_X$ of +200Oe, and then the heterostructure is also initialized to -$m_z$. If $I_A$ or $I_B$ is 1, making $I_A + I_B > I_c$, which leads to the deterministic switching from -$m_z$ to +$m_z$. "NOR", "NAND" and "NOT" can be realized by changing the polarity of $H_X$ and the polarization field of BFO.

In this work, the PMA and SOT-induced magnetization switching have been successfully realized in the BFO-based heterostructures. The critical current for the SOT-induced magnetization switching has been changed by nearly 300% with the opposite polarization fields of the BFO films. The effective SHA of the heterostructure with the up-ward polarized BFO film is 2.72 times larger than the one with the down-ward polarized BFO film, which is due to different oxygen vacancies distribution in the BFO films. We also demonstrated that our heterostructures have great potential to be used in future memory and programmable logic devices. Accordingly, this work supplies a new way to manipulate SOT through changing polarization field, which may open a fascinating direction combining spintronics and multiferroic materials.

**Notes**

The authors declare no competing financial interest.

**SUPPORTING INFORMATION:**

The Supporting Information is available free of charge on the ACS Publications website.

Details of our Hall bar's fabrication process; Out-of-plane Piezoelectric force microscopy images of STO/BFO and STO/SRO/BFO; The hysteresis loop and AHE measurements of BFO hetero and S-BFO hetero; Dependence of the first $V\omega$ and second $V2\omega$ harmonic Hall voltages on in-plane magnetic field (X-direction and Y-direction) of the measurement current; The first $V_\omega$ and second $V_{2\omega}$ harmonic Hall voltages measured under small in-plane magnetic fields; Angular dependence of $R_H$ and correlation calculation; Co-$L_{2,3}$ XAS spectra of BFO hetero and S-BFO hetero;

**Figures captions**

Figure 1 (a) AFM image of the SRO/BFO bilayer. (b) XRD pattern of the SRO/BFO bilayer. (c) Schematic diagram of the Hall bar structure. (d) The RSM in (103) orientation for the SRO/BFO bilayer.

Figure 2 R-I curves of (a, b) BFO hetero and (c, d) S-BFO hetero, under in-plane magnetic fields. For BFO hetero, $\Delta R=3.4\,\Omega$, while for S-BFO hetero, $\Delta R=0.4\Omega$.

Figure 3 The first $V_\omega$ and second $V_{2\omega}$ harmonic Hall voltages measured under in-plane magnetic field for BFO hetero (red curves) and S-BFO hetero (green curves). The measurement current is 2.1 mA with the frequency of 111.2 Hz.

Figure 4 Illustrations of the SOT effect in BFO hetero and S-BFO hetero. The surface of BFO can be adjusted into Oxygen-vacancy-poor (a) or Oxygen-vacancy-rich (b) states with different directions of spontaneous polarization fields. Insets in (c)-(f) are the amplified parts of corresponding R-I curves denoting critical currents.

Figure 5 (a) A schematic diagram of memory devices comprising of the multiferroic BFO-based heterostructures. (b) Truth tables of the "AND" gate and "OR" gate. Three numbers are used for recording our logic outputs and inputs. The first numbers (red ones) present the final output in logic operations. The second and third numbers present the Inputs from $I_A$ and $I_B$ respectively.

Figure 1:

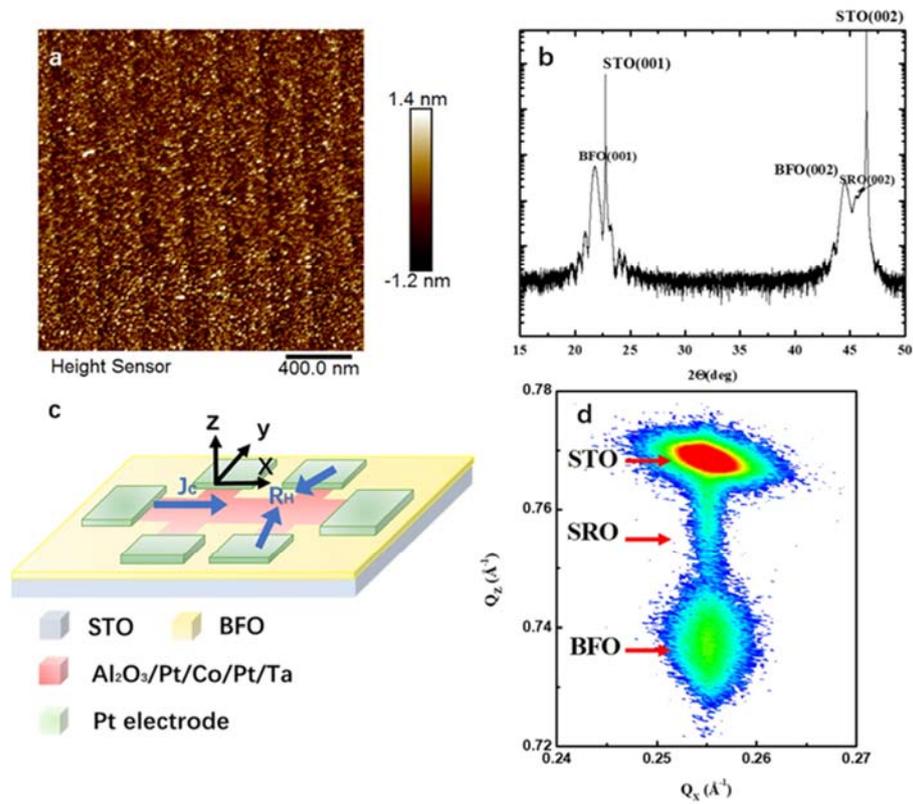

Figure 1 (a) AFM image of the SRO/BFO bilayer. (b) XRD pattern of the SRO/BFO bilayer. (c) Schematic diagram of the Hall bar structure. (d) The RSM in (103) orientation for the SRO/BFO bilayer.

Figure 2:

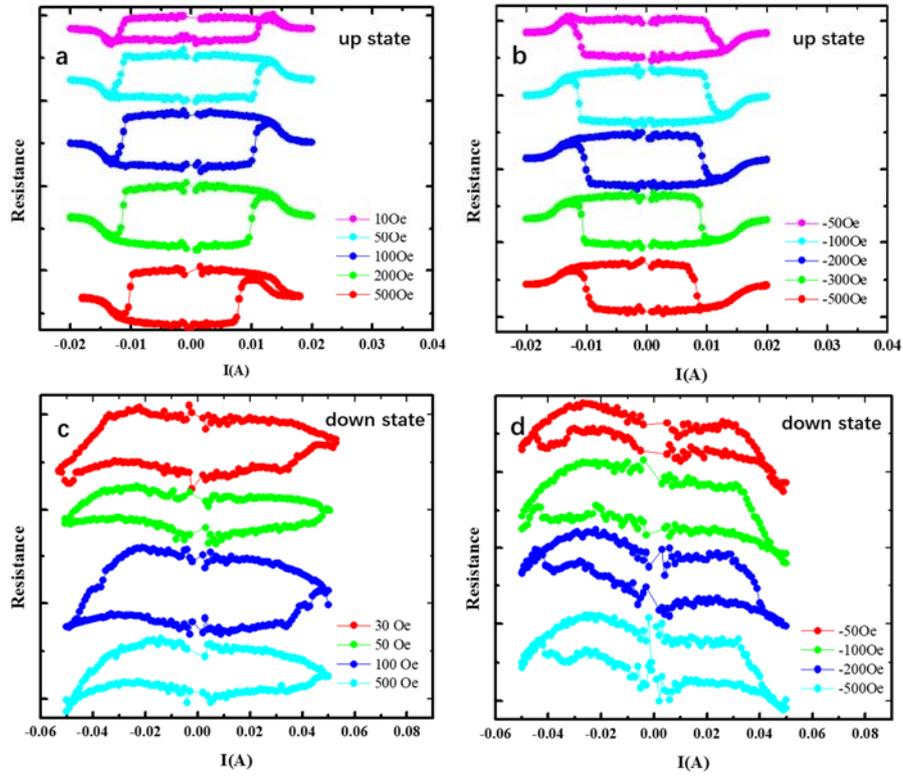

Figure 2 R-I curves of (a, b) BFO hetero and (c, d) S-BFO hetero, under different in-plane magnetic fields. For BFO hetero, $\Delta R=3.4 \Omega$, while for S-BFO hetero, $\Delta R=0.4\Omega$.

Figure 3:

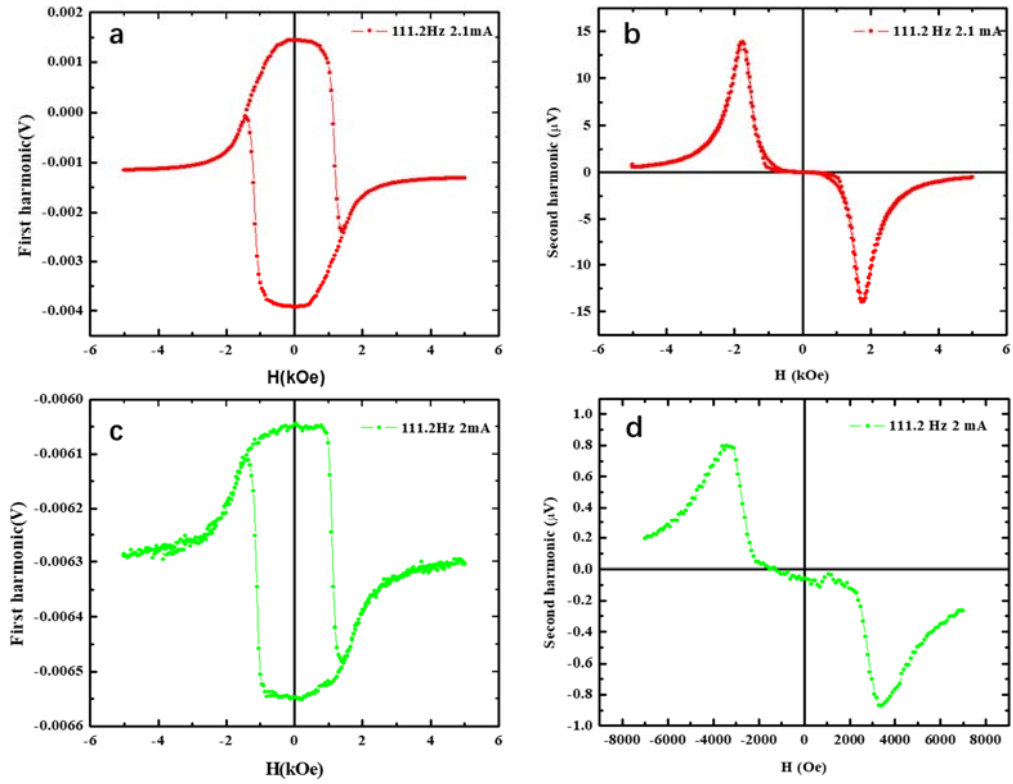

Figure 3 The first $V_\omega$ and second $V_{2\omega}$ harmonic Hall voltages measured under in-plane magnetic field for BFO hetero (red curves) and S-BFO hetero (green curves). The measurement current is 2.1 mA with the frequency of 111.2 Hz.

Figure 4:

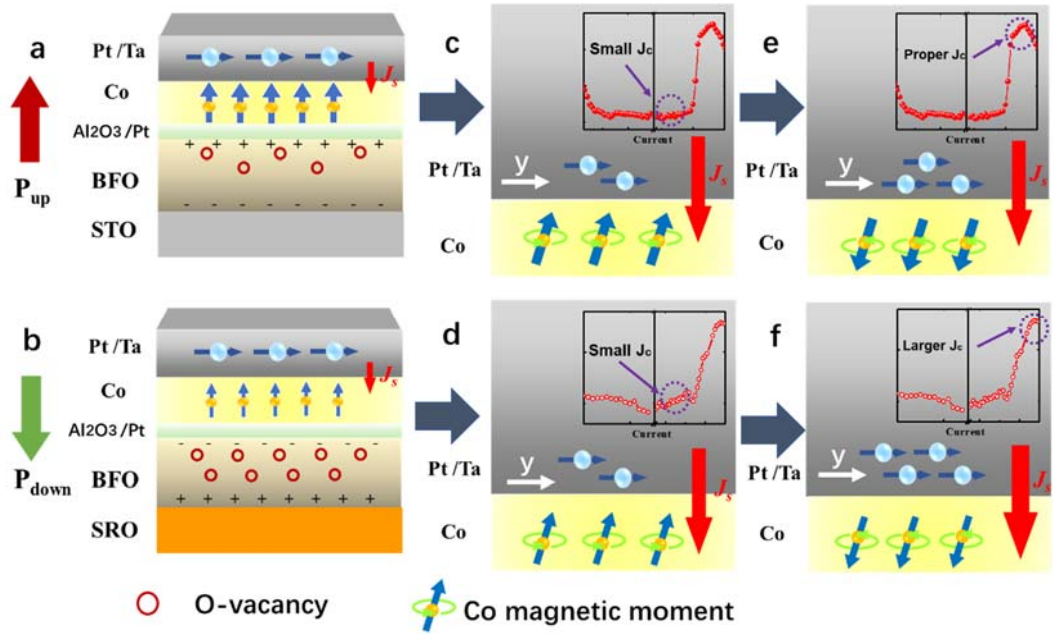

Figure 4 Illustrations of the SOT effect in BFO hetero and S-BFO hetero. The surface of BFO can be adjusted into Oxygen-vacancy-poor (a) or Oxygen-vacancy-rich (b) states with different directions of spontaneous polarization fields. Insets in (c)-(f) are the amplified parts of corresponding R-I curves denoting critical currents.

Figure 5

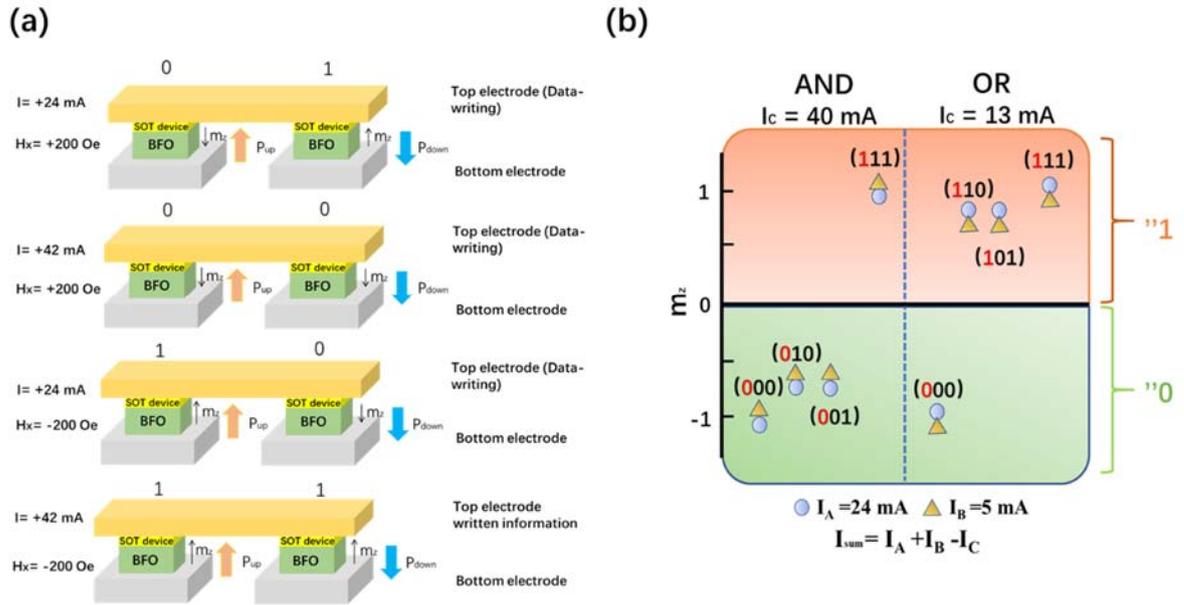

Figure5 (a) A schematic diagram of memory devices comprising of the multiferroic BFO-based heterostructures. (b) Truth tables of the "AND" gate and "OR" gate. Three numbers are used for recording our logic outputs and inputs. The first numbers (red ones) present the final output in logic operations. The second and third numbers present the Inputs from $I_A$ and $I_B$ respectively.